\documentstyle[pre,aps,epsfig,floats]{revtex}

\begin{document}

\draft

\title{Bond-orientational Order in Melting of Colloidal Crystals\footnote[1]{Journal of the Korean Physical Society 49, 1682(2006)}}

\author{Xin Qi, Yong Chen,\footnote[2]{Author to whom correspondence should be addressed. Email address: ychen@lzu.edu.cn} Yan Jin, and Yao-Hui Yang}

\address{Institute of Theoretical Physics, Lanzhou University, Lanzhou 730000}

\date{\today}

\maketitle

\begin{abstract}
Using Brownian dynamics simulation, we study the orientational order in the melting transition of a colloidal system with a $'$soft$'$ Yukawa potential. The bond-orientational order parameter $\Phi_{6}$ and the bond-orientational order function $g_B(r)$ are calculated in two-dimensional systems. We found that a two-stage transition and a hexatic phase indeed exist in two-dimensional melting, which is consistent with the prediction of the Kosterlitz-Thouless-Halperin-Nelson-Young theory. For comparing with the melting process in three-dimensional systems, we introduce the probability distribution of the single-particle local-order parameter. Based on extensive simulations, the breakdown of local order is qualitatively suggested only to occur on the fractional part of the colloidal system for the two-dimensional melting, but in three-dimensional melting, that breakdown takes place on the whole system at the same time.
\end{abstract}

The melting transition in two-dimensional (2D) systems has received much attention during the past ten years. Unlike higher-dimensional systems, 2D solids possess a quasi-long-range positional order. This property implies that the melting transition in two dimensions is different from that in 3D systems. In $1973$, Kosterlitz and Thouless suggested that the melting transition in 2D crystals was a continuous process mediated by the dissociation of dislocation pairs {\cite {1}}. Later, the resulting phase of this continuous transition was shown not to be an isotropic phase because it still had a quasi-long-range orientational order {\cite {2}}. In $1979$, Young pointed out that a second-transition, which was induced by the formation of disclinations, would drive this so-called hexatic phase into liquid {\cite {3}}. This theory, which is well-known as the Kosterlitz-Thouless-Halperin-Nelson-Young (KTHNY) theory {\cite {4}}, thus proposes the following two-stage scenario for melting: The solid first undergoes a continuous transition to become a hexatic phase with a quasi-long-range orientational order; then, another continuous transition drives the hexatic phase to a disorder liquid.

During the years, a large number of experiments and simulations have been performed to verify the KTHNY theory {\cite {5,6,7,8}}; however, results remain controversial, which It may be caused by the size effect of the systems. because long-wavelength fluctuations play an important role in the KTHNY theory, they are cut off in finite-sized simulations. In particular, very large-scale simulations of Lennard-Jones systems seem to provide evidence for the existence of the hexatic phase {\cite {9}}. However it was argued that this transition might depend on the specific properties, such as the interparticle potential of the studied systems {\cite {10,11}}. Recently, a series of experiments was performed to calculate the elastic moduli and the dislocation interaction of 2D colloidal crystals {\cite {12,13,14,30}}. The renormalized Young's modulus of the crystals, $K_R$, was found to be consistent with the KTHNY theory, and the dissociation of dislocations was observed experimentally.

This work is devoted to a study of the melting transitions of charged colloidal crystals. Colloidal crystals are known to be ideal model systems for statistical physics. A particular advantage of such systems is that, due to the particle size, the individual colloid positions and motions can be directly observed. Moreover, the inter-particle potentials in such colloidal crystals are in most cases precisely known. For example, the well-known Dejaguin-Landu-Verwey-Overbeek (DLVO) potential gives a good description for the effective pair interaction of one-component model of the colloidal systems. However, for a dilute charged-stabilized colloidal system in which many-body interactions can be ignored, a pairwise Yukawa potential, which is only the electrostatic part of the DLVO potential, is suitable {\cite {15,24,25,26}}:
\begin{equation}
V\left( r\right) =U_{0}\sigma\exp[-\lambda(r-\sigma)/\sigma]/r,   \label{eq-1}
\end{equation}
where $U_{0}$ sets the energy and $\sigma$ the length scale. The screening parameter $\lambda$ describes the $'$softening$'$ of the particles: When $\lambda$ increases from zero to infinity, the cores of interacting particles change from very soft to extremely hard {\cite {16}}.

While most of the previous research focused on hard-core colloidal particles, we study a $'$soft$'$ disk system with Yukawa interactions. However, no consensus was achieved even in such systems {\cite {10,27}}. In this paper, the bond-orientational order parameter $\Phi_{6}$, the probability distribution of the single-particle order parameter $P(\phi_{6})$, and the orientational correlation function $g_{B}(r)$ are calculated to describe the melting processes of 2D $'$soft$'$ colloidal crystals. The translational order is ignored because it cannot describe the properties of a two-stage melting transition. Melting in 3D systems was also studied for comparison with 2D systems.

First, we briefly describe the standard Brownian-dynamics simulation method {\cite {16,17,18}}. It is based on a finite-difference integration of the irreversible Langevin equations. If the effect of the hydrodynamic interactions is ignored, the equations of motion of $N$ particles at equilibrium are derived from Eq. (\ref{eq-1}) as
\begin{equation}
\xi \dot{\mathbf{r}}_{i}(t) = {\mathbf{F}}_{i} (t)+ {\mathbf{R}} (t),
\label{eq-2}
\end{equation}
where $i = 1,....N$ labels the $N$ particles, $\xi$ is the friction coefficient, ${\mathbf{R}}(t)$ is the Langvein random force of the solvent, and ${\mathbf{F}}_{i}$ is the total inter-particle force on particle $i$. The finite difference integration of Eq. (\ref{eq-2}) is
\begin{equation}
{\mathbf{r}}_{i}(t +\triangle t) = {\mathbf{r}}_{i}(t)+(1/\xi){\mathbf{F}}_{i}(t)\triangle t +({\mathbf{\triangle r}})_{R} + O((\triangle t)^{2}),  \label{eq-3}
\end{equation}
where $({\mathbf{\triangle r}})_{R}$ is a random displacement sampled from a Gaussian distribution of zero mean and variance
\begin{equation}
\overline{(\triangle {\mathbf{r}})^{2}_{R}}=4D_{0}\triangle t.   \label{eq-4}
\end{equation}
$D_{0}=k_{B}T/\xi$ is the short-time diffusion coefficient, where $k_{B}$ is the Boltzmann constant and $T$ is the temperature.

In all simulations, we used reduced units such that $\sigma = 1$, $U_{0} = 1$, and $\rho = N /V = 1$. The screening parameter $\lambda = 8$. In the case of 2D simulation, we used a periodically repeated rectangular simulation box of volume $V$ with $N = 2500$ particles and started from a triangular lattice; The cutoff $r_c$ was set as $4.1\bar{r}$, where $\bar{r}=\sqrt{V/N} = {\rho}^{-1/2}$. The scale of the simulation box was set to $x:y=\sqrt{3}:2$ in order to minimize the effect of the boundary. Particle trajectories were generated according to Eq. (\ref{eq-3}), and the simulation data were gathered within the range of $5\tau_{B}$ ($\tau_{B}= \sigma^{2}\xi/U_{0}$) after a long equilibration time (more than $45\tau_{B}$). In our simulations, we varied the reduced temperature $T^{*}$ while the other parameters $\sigma$, $U_{0}$, $\rho$, and $\lambda$ were fixed. Several different runs corresponding to different reduced temperatures were done.

There are some remarks to be made on the simulation details. First, the timestep $\triangle t$ depends on the screening parameter $\lambda$. When the potential becomes $'$hard$'$, a shorter $\triangle t$ must be chosen. Another important thing is the relaxation time of the 2D system. It should be very long so that equilibrium will be reached if a two-stage continuous melting transition occurs. In particular, the $'$hysteresis loops$'$ will be incorrect when the system is heated or cooled too fast {\cite {19}}. Lastly, the center of mass of the system was fixed during the simulation in order to avoid super drift.

Melting in 3D systems was also simulated for comparison with 2D melting. The method is the same as in 2D systems except that the $4$ on the right-hand side of Eq. (\ref{eq-4}) is changed to $6$ and $\bar{r} = \rho ^{-1/3}$ in the cutoff length ${r}_c = 4.1\bar{r}$. We used a system of $500$ particles with periodic boundary conditions and started from FCC structures.

The bond-orientational function $g_{B}(r)$ is used to identify the existence of the hexatic phase {\cite {20}}. It is defined as
\begin{eqnarray}
g_{B}(r) = \displaystyle\frac{\left\langle \psi_{6}^{*}({\mathbf{r}}')\psi_{6}({\mathbf {r'-r}}) \right\rangle}{\left\langle \delta(\mathbf{r})\delta(\mathbf{r'-r})\right\rangle}\nonumber = \frac{\displaystyle \left\langle \sum _{i} \sum_{i<j}\psi_{6}({\mathbf{r_{i}}})\psi_{6}({\mathbf{r_{j}}}) \delta(\mathbf{r}-\mathbf{r_{ij}}) \right\rangle}{\displaystyle g(r)},    \label{eq-5}
\end{eqnarray}
where
$g(r)=\langle\delta(\mathbf{r}')\delta(\mathbf{r}'-\mathbf{r})\rangle$ is the pair distribution function and $\psi_{6}(\mathbf{r})$ is the local bond orientational order parameter and is given by
\begin{equation}
\psi_{6}({\mathbf{r}}_m)=\frac{1}{N_b}\displaystyle \sum^{N_b}_{n=1}e^{6i\theta_{mn}}.
\label{eq-6}
\end{equation}
Here, $N_b$ denotes the number of nearest neighbors of the $m$th particle, and $\theta_{mn}$ is the angle between some fixed axis and the bond joining the $m$th particle with an $n$th particle. According to the KTHNY theory, the bond-orientational function $g_B(r)$ will have an algebraic decay in a hexatic phase and an exponential decay in the liquid phase.

The order parameter $\Phi_6$ was introduced by Nelson and Halperin to characterize the structural order in 2D systems {\cite {2}}. It is given by {\cite {21}}
\begin{equation}
\Phi_6=\displaystyle \left\langle \frac{1}{N}\sum^{N}_{m=1} \frac{1}{N_b}\sum_{n=1}^{N_b}e^{6i\theta_{ mn}} \right\rangle.
\label{eq-7}
\end{equation}
When the system belongs to the fluid phase, $|\Phi_6|^2\ll 1$. On the other hand, $|\Phi_6|^2\sim1$ means that the system is a perfect crystal. Especially, we define a single-particle order parameter $\phi_6$, which is the ensemble average of the local bond-orientational parameter $\psi_6$:
\begin{equation}
\phi_6=\displaystyle \left\langle  \frac{1}{N_b}\sum_{n=1}^{N_b}e^{6i\theta_{ mn}} \right\rangle.
\label{eq-8}
\end{equation}
Clearly, $\phi_6$ is the order parameter of a single particle, and one can calculate $\phi_6$ for each particle in our simulation box.

As $\Phi_6$ in two dimensions, the bond order parameter $Q_6$ is used to describe the orientational order of a 3D system. It is defined by calculating the spherical harmonics over all bonds {\cite {22}}. Value of $Q_6$ from $0.354$ to $0.663$ represent systems with crystal structures {\cite {23}}. For example, when the system is a perfect FCC crystal, $Q_6=0.57452$, and when system is a liquid, $Q_6\ll 1$. As $\phi_6$ in 2D systems, we also introduce a single-particle local order parameter $q_6$, which is the order parameter of a single particle. The probability distributions of $q_6$ and $\phi_6$ represent a possible way to qualitatively study 2D and 3D melting transitions, as shown in the following context.

\begin{figure}[h]
\begin{center}
\epsfig{figure=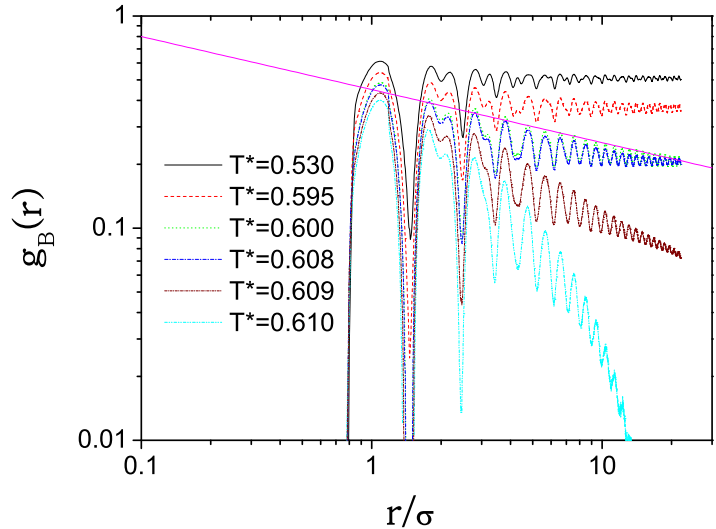,width=11cm}
\caption{Bond-orientational function $g_B(r)$ versus $r/ \sigma$ for various values of the reduced temperature $T^*=k_BT/U_0$. The 2D simulation system consists of $2500$ particles and the screening parameter $\lambda = 8$. The straight line, with slope $-1/4$, is a guide for the eyes.}
\label{Fig1}
\end{center}
\end{figure}

The 2D simulation results for the bond-orientational functions $g_B(r)$ are given in Fig. \ref{Fig1} with different reduced temperatures $T^*=k_B T/U_0$. At the low reduced temperature, the bond-orientational order function $g_B(r)$ does not decay, which implies the system is in a solid phase with a long-range bond-orientational order, but when the reduced temperature rises to $0.600$, $g_B(r)$ decays algebraically with an exponent less than $-1/4$, the upper limit of the existence of a stable hexatic phase predicted by the KTNHY theory. This provides very strong evidence for the hexatic phase. As the temperature keeps increasing, the system becomes a disordered liquid, and $g_B(r)$ decays exponentially. The exponent of the algebraic decay always decreases with increasing temperatures in the hexatic phase. However, this decrease of the exponent is very small because the temperature region of the haexatic phase is only $0.008$. As a result, the temperature dependence is hardly observed in the simulation results. It should be noted, in Fig. \ref{Fig1}, that $g_B(r)$ becomes flat for larger $r$ due to the finite-size effect, such as $g_B(r)$ at the hexatic phase for $r > 20.00$. In Fig. \ref{Fig1}, we only present $g_B(r)$ for $r \leq 22.00$. The length of the x-axis of our simulation box is $55.836 \times \sigma$.

Note that because of the large fluctuation near the critical point, the phase boundaries we obtained are not very precise, but what we are interested is the process of the melting transition. Additionally, we perform extensive simulations in very large ranges of $T^{*}$ from $0.050$ to $1.200$ and of $\rho$ from $0.6$ to $1.2$. Moreover, a two-stage melting transition is clearly observed in each case. Thus,we can conclude that melting transition of a soft disk system is absolutely a solid-hexatic-liquid process.

It is worthwhile to note that the $g_B(r)$ decays algebraically with an exponent is $-2/3$ at $T^* = 0.609$, which is faster than the $-1/4$, predicted by the KTHNY theory. In our simulations at this reduced temperature, it took a very long time for the systems to reach equilibrium. We conjecture that $T^* = 0.609$ is the critical reduced temperature for the transition from the hexatic phase to a disordered phase. The detailed properties of the system at $T^* = 0.609$ or the hexatic-disordered transition are still unclear and need to be studied further.

\begin{figure}[h]
\begin{center}
\epsfig{figure=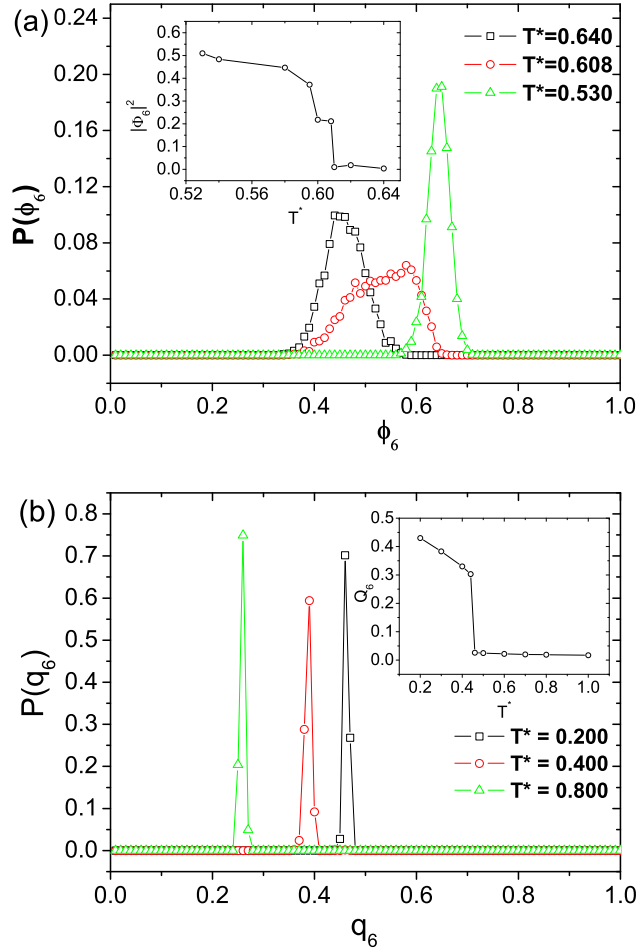,width=11cm} \caption{(a) Probability distributions of the single-particle order parameter $\phi_6$ in 2D melting.  $P(\phi _6)$ exhibit an orderless distribution at $T^*=0.608$, which implies the existence of a hexatic phase at this temperature. The inset shows the value of the order parameter $|\Phi_6|^2$ at different $T^*$. Two drops occurs, one at $T^{*}=0.600$ and the other at $T^{*}=0.608$, implying a two-stage transition. (b) Probability distributions of $q_6$ in 3D systems. The values are distributed over a very narrow range. The inset shows the value of $Q_6$ at different $T^{*}$. The collapse at $T^{*}=0.440$ represents the occurrence of a first-order transition.} \label{Fig2}
\end{center}
\end{figure}

In Fig. \ref{Fig2}, we present the simulation results for the bond-orientational order parameters in melting of 2D and 3D colloidal crystals. At low temperature, $|\Phi_6|^2$ has a large value, which implies that the 2D system is in a solid phase [see the inset in Fig. \ref{Fig2} (a)], and it decreases slowly with increasing $T^*$. The first drop is found at $T^*=0.600$, corresponding to the fact that a transition occurs at this temperature. Then, $|\Phi_6|^2$ returns to a slow decrease until the second drop takes place at $T^*=0.608$. We suggest that a two-stage phase transition, indeed, occurs in the melting of 2D systems with soft Yukawa potentials. In contrast, the single collapse of $Q_6$ at $T^*=0.440$, as shown in the inset of Fig. \ref{Fig2}(b), implies that 3D melting with a soft Yukawa interaction is just a well-understood first-order transition. Obviously, the bond-orientational order parameter is very sensitive to the structure transition of the simulation systems, which can be used to confirm the emergence of two-stage melting.

The probability distribution of the single-particle order parameter $P(\phi_6)$, as shown in Fig. \ref{Fig2}(a), exhibits a Gaussian-type behavior of both high and low temperatures, and the variance increases with temperature. However, when the system is at an intermediate $T^*$, $P(\phi_6)$ is not a Gaussian-like curve anymore. We conjecture that the orderless distribution in Fig. \ref{Fig2}(a) is caused by the appearance of isolated disclinations ($5-$ and $7-$fold coordinated particles having exclusively $6$-fold coordinated neighbors), which is characteristic of the hexatic phase destroying the translational symmetry of the crystal, as observed in Ref. {\cite13}. In the case of 3D systems, $P(q_6)$ keeps a Gaussian-like distribution for the entire melting process [see Fig. \ref{Fig2}(b)]. This appears to be due to an enhancement of the thermal motions of all the particles with increasing temperatures. Also the breakdown of the ordered phase takes place for the whole systems at the same time. Moreover, it is worth mentioning that, in all the simulation results, the variance of the distribution of melting in a 3D system is obviously smaller than that in a 2D system. Additionally, we suggest that $P(\phi_6)$ is more intuitive and easier than $g_B(r)$ for presenting detailed structural information in the melting process.

In conclusion, we have performed Brownian dynamics simulations on the melting of 2D colloidal crystals in which particles interact with a $'$soft$'$ Yukawa potential. A two-stage melting transition and a metastable hexatic phase between the solid and the liquid phases are observed in our simulations, which are consistent with the KTHNY theory. On the other hand, the melting in 3D soft Yukawa systems is just a one-step phase transition. In addition, through simulations of the bond-orientational order parameter and the probability distribution of the single-particle local-order parameter, the 3D melting is thought to occur with the breakdown of the locally ordered phase over the entire system at the same time.

\begin{acknowledgments}
This work was partially supported by the SRF for ROCS, SEM, and by the Interdisciplinary Innovation Research Fund For Young Scholars, Lanzhou University.
\end{acknowledgments}

\end{document}